\documentclass[12pt]{article}

\advance \textheight by \topskip
\def\R{\mathbb R}

\def\N{\mathbb N}
\def\Z{\mathbb Z}

\def\asl{\mathfrak{sl}}

\def\pd{\partial}

\def\mop #1{\mathop{\sf #1}\nolimits}

\usepackage{amsmath,amssymb,epsfig}

\begin{document}
\title{
{\bf Additional restrictions on quasi-exactly solvable
systems}}
\author{{\sf Sergey Klishevich}\thanks{E-mail:
klishevich@ihep.ru}\\
{\small {\it Institute for High Energy Physics, Protvino,
Russia}}}
\date{}
\maketitle

\vskip-1.0cm

\begin{abstract}
In this paper we discuss constraints on two-dimensional
quantum-mechanical systems living in domains with boundaries. The
constrains result from the requirement of hermicity of corresponding
Hamiltonians. We construct new two-dimensional families of formally
exactly solvable systems and applying such constraints show that in real
the systems are quasi-exactly solvable at best. Nevertheless in the
context of pseudo-Hermitian Hamiltonians some of the constructed
families are exactly solvable.
\end{abstract}
\newpage

\section{Introduction}

It is well known that exactly solvable systems play very important role
in quantum theory. Unfortunately number of such systems is quite
limited. This considerably narrows their applications. Such a situation
stimulates interest to quasi-exactly solvable systems
\cite{qes, shifman, ushv1, ushv2}. In contrast to exactly solvable
models in quasi-exactly solvable systems the spectral problem can be
solved partially. Nevertheless such systems are very interesting.
Besides modeling physical systems~\cite{appl} they can be used as an
initial point of the perturbation theory or to investigate various
nonperturbative effects~\cite{Aoyama}. Furthermore, recently in the
series of papers~\cite{nsusy} (see also Refs.~\cite{andrian2, klish04})
it was revealed a connection between quasi-exactly solvable models and
supersymmetric systems with polynomial superalgebras~\cite{andrian}.
Also we can hope that progress in understanding quasi-exactly
solvable quantum-mechanical systems will allow to find out methods of
constructing quasi-exactly solvable models in quantum field theory.

The paper has the following structure. In section~\ref{general} a brief
introduction into the Lie-algebraic approach to constructing
quasi-exactly solvable systems is given. In section~\ref{bounds} we
discuss constraints on wave functions of systems living in domains with
boundaries. It is shown that for corresponding Hamiltonians to be
Hermitian the wave functions must have a specific behaviour at
boundaries. The role of the constraints is illustrated by an example of
a known two-dimensional quasi-exactly solvable system. In
section~\ref{new} we construct new families of formally exactly solvable
systems. Application of the restriction on the behaviour of wave
functions at boundaries leads to the conclusion that the systems are
quasi-exactly solvable even if the wave functions are normalizable.
Discussion of results is presented in section~\ref{end}.

\section{General aspects of quasi-exactly solvable systems}
\label{general}

From the general viewpoint an operator is quasi-exactly solvable if it
has a finite-dimensional invariant subspace.
Bearing in mind applications to the quantum mechanics we are interested
in second order differential operators, Hamiltonians. The most famous
method of constructing quasi-exactly solvable differential operators
is the Lie-algebraic approach \cite{qes,shifman,ushv1,ushv2}. In this
section we briefly discuss of main ideas of the approach.

The one-dimensional and multidimensional cases have to be treated
separately. The one-dimensional case is the most elaborated one. The
simplest finite-dimensional subspaces are spaces which admit the
following monomial basis:
$$
 \mathcal F_n=\mop{span}\left\{1, z,z^2,\ldots,z^n\right\}.
$$
For such spaces it is not difficult to find invariant (quasi-exactly
solvable) differential operators of the first order:
\begin{align}\label{sl2}
 J_-&=\frac d{dz}, &
 J_0&=N_z-\frac{n-1}2,&
 J_+&=z\left(N_z-n\right),
\end{align}
where $N_z=z\frac d{dz}$. These operators form finite-dimensional
representations of the algebra $\asl(2,\R)$ for $n\in\N$. Any second
order operator taken as quadratic combination of the first order
operators,
\begin{equation}\label{Hsl2}
  H'=C^{ab}J_aJ_b+C^aJ_a ={} - P_4(z)\frac{d^2}{dz^2}+\ldots,
\end{equation}
is automatically quasi-exactly solvable\footnote{In principle one has to
proof that the finite-dimensional subspace corresponds to wave functions
with finite norms. In the case of the algebra $\asl(2,\R)$ this question
was completely investigated in Ref. \cite{kamran93}.}. Here $P_4(z)$ is
a polynomial of order 4. In the one-dimensional case any second order
operator can be reduced to the Schr\"odinger form:
\begin{align*}
 \Psi(x)=e^{a(z)}P(z)\quad&\text{with}\quad
    x=\pm\int\frac{dz}{\sqrt{P_4(z)}}&\Rightarrow&&
    H&={}-\frac{d^2}{dx^2}+V(x),
\end{align*}
where $\Psi(x)$ is a wave function of the Hamiltonian in the canonical
form. This is essential difference of the one-dimensional case from
multidimensional systems.

The complete classification of linear differential operators with
invariant subspaces of monomials is given in Ref.~\cite{turb&post}
(more recent discussions see in Ref.~\cite{kamran04}).

In multidimensional case the situation is much more complicated
\cite{shifman}. Nevertheless the idea is the same. We construct
finite-dimensional representation of some Lie algebra in terms of first
order differential operators \cite{shifman, kamran91}, then one can look
for Hamiltonians considering quadratic combinations of the first order
operators:
\begin{align*}
  H'&=C^{ab}J_aJ_b+C^aJ_a ={}
  - g^{\mu\nu}\left(\nabla_\mu-A_\mu\right)
        \left(\nabla_\nu-A_\nu\right) + V,
\end{align*}
with
\begin{equation*}
 V=g^{\mu\nu}A_\mu A_\nu - A^\mu{}_{;\mu},
\end{equation*}
where $\nabla_\mu$ and "$;\mu$" stand for the covariant derivative
corresponding to the metric $g_{\mu\nu}$ and Such a Hamiltonian can be
reduced to the Schr\"odinger form only if the self-consistency
conditions are satisfied:
\begin{equation}\label{dA}
 \pd_\mu A_\nu - \pd_\nu A_\mu = 0.
\end{equation}
From another point of view these equations are the necessary condition
for the Hamiltonian to be Hermitian. The general solution of the
constraints is unknown! Therefore in the multidimensional case the
general classification of quasi-exactly solvable linear differential
operators is unknown even for invariant subspaces of monomials. However
as it was pointed out in Ref.~\cite{ushv1} the problem with resolving
the constraints \eqref{dA} can be avoided if one passes from $D$- to
$(D+1)$-dimensional system. Details of the procedure can be found in
Ref.~\cite{ushv1}). Nevertheless here we do not adopt this standpoint.

\section{Boundary conditions}\label{bounds}

When constructing quasi-exactly solvable operators in the Lie-algebraic
approach very often resulting systems live in domains with
boundaries.\footnote{Here we imply the multidimensional case.} If so one
has to take into account behaviour of wave functions at the boundaries
in addition to their normalizability.

Generally the system must evolve in a domain with a positively defined
metrics
\begin{equation}\label{S}
  S=\left\{x^\mu\,\big|\,g_{11}>0\cup\mop{det}\|g_{\mu\nu}\|>0
   \right\}.
\end{equation}
Here we investigate situation when components of the inverse metric
$g^{\mu\nu}$ are polynomial. Therefore it is more convenient to define
the boundaries of the domain \eqref{S} as roots of the inverse metric
determinant:
$$
 \pd S\subset\left\{x^\mu\,\big|\,\mop{det}\|g^{\mu\nu}\|=0\right\}.
$$
To provide hermicity of the corresponding Hamiltonian the following
conditions have to be implied:
\begin{equation}\label{herm}
   \sqrt gg^{\mu\nu}\varphi\pd_{\nu}\psi\Big|_{\pd S}=0
\end{equation}
for any $\varphi(x)$ and $\psi(x)$ from the domain of the Hamiltonian.

In the two-dimensional case the boundary can be locally given by the
equation $x=\xi(y)$. Then from the conditions \eqref{herm} one can infer
that the wave functions from the domain of the Hamiltonian have the
following behaviour at the boundary:
\begin{equation}\label{bound}
 \psi\sim g^{-\frac 14}\left|x-\xi(y)\right|^\alpha
 \quad\text{with}\quad\alpha>\frac 12.
\end{equation}
The normalizability of such a wave function leads to the inequality
$\alpha>-\frac 12$. So the hermicity implies the more strict
inequality but often it is not taken into account. For example,
consider the following Hamiltonian \cite{shifman}:
\begin{align}\label{Hex}
 {}-H'&=x\left(1+x\right)\pd^2_x + y\left(1+y\right)\pd^2_y
   - 2xy\pd^2_{xy}\notag\\&
    +\left(1+x\right)\left(1-cx\right)\pd_x
    +\left(1+y\right)\left(1-cy\right)\pd_y
    +2c\left(jx+\tilde\jmath y\right).
\end{align}
It can be represented in terms of generators of the algebra
$\asl(2,\R)\oplus\asl(2,\R)$. In this case the inverse metric is
$$
 \left\|g^{\mu\nu}\right\| =
  \begin{pmatrix}
   x\left(1+x\right)&-xy\\-xy&y\left(1+y\right)
  \end{pmatrix}.
$$
Its determinant has the form
$$
 \mop{det}\|g^{\mu\nu}\|=xy\left(1+x+y\right).
$$
Therefore we can consider the system in the domain
$S=\left\{(x,y)\,\big|\,x>0 \cup y>0\right\}$. In principle there
are four domains with positively defined metric but the consequence is
the same for all of them.
The wave functions corresponding to the quasi-exactly solvable sector
have the structure
$$
 \psi=g^{-\frac 14}e^{c(xy+x+y)}Pol(x,y),
$$
where $Pol(x,y)$ is a polynomial in $x$ and $y$.
One can see that such functions are normalizable in the domain $S$ for
$c<0$. However functions of such a form do not belong to the domain
of the Hamiltonian since they have improper behaviour at the boundaries.
Therefore, actually the Hamiltonian \eqref{Hex} is not quasi-exactly
solvable. Nevertheless it can be of interest in the context of
pseudo-Hermitian Hamiltonians \cite{psH} (also see the discussion
below).

\section{New families of quasi-exactly solvable systems}
\label{new}
\subsection*{Formally exactly-solvable systems}

Exactly solvable systems form a subset of quasi-exactly solvable
systems, because evidently they have (an infinite flag of)
finite-dimensional invariant subspaces.

Let us start discuss of this point in the context of the Lie-algebraic
approach from the one-dimensional case \eqref{sl2}, \eqref{Hsl2}. One
can select two exactly solvable operators of the first order,
$\frac d{dz}$ and $z\frac d{dz}$. A Hamiltonian constructed in terms of
these operators has the following general form:
\begin{align*}
  H'&={}-P_2(z)\frac{d^2}{dz^2} + P_1(z)\frac d{dz}.
\end{align*}
The resulting Hamiltonian is exactly-solvable, but only formally. To
proof the exact solvability one has to check normalizability of
corresponding wave functions. Not all systems given by the Hamiltonian
are exactly solvable in this sense.

Now we pass to the two-dimensional case. The following first order
operators
\begin{align}\label{rF2}
 N_y&=y\pd_y, & N_x&=x\pd_x, & L_0=&\pd_y, & L_p&=y^p\pd_x,
\end{align}
where $p=0,\ 1,\ \ldots,\ m$, are exactly solvable because they are
invariant on
the spaces
\begin{align}\label{F2}
 \mathcal F_{m,n}&=\mathop{\sf span}_
   {\genfrac{}{}{0pt}{}{ma+b\leqslant mn}{a,b\in\Z_+}}
   \left\{x^ay^b\right\}
\end{align}
for any $n\in\N$. For any fixed $n$ it is possible introduce yet another
operator
$$
 L_{m,n}=y\left(mN_x+N_y-mn\right),
$$
which is quasi-exactly solvable. All these operators form a series of
representations of the algebra
$\R^{m+1}\mathbin{\lefteqn{\hskip 0.8 mm +}\supset}\mathfrak{gl}(2,\R)$
\cite{kamran91}. It is worth noting that the case $m=1$ has to be
considered separately since in this case there exist additional
quasi-exactly solvable operators of the first order, e.g. see
Ref~\cite{kamran99}.

The general form of the Hamiltonian constructed in terms of the first
order operators \eqref{rF2} is
\begin{align}
 {}-H&=\left(p_0x^2 + P_m(y)x + P_{2m}(y)\right)\pd_x^2
     + \bigl(P_1(y)x + P_{m+1}(y)\bigr)\pd_{xy}^2
     + P_2(y)\pd_y^2
     \notag\\[-3mm]\label{H2}\\[-3mm]\notag
    &+ \bigl(q_0x + Q_m(y)\bigr)\pd_x + Q_1(y)\pd_y,
\end{align}
where $p_0,\ q_0\in\R$, $P_q(y)$ and $Q_p(y)$ are polynomials of degree
$q$ and $p$ with real coefficients.
The Hamiltonian \eqref{H2} does not depend on the integer parameter $n$
and, as a consequence, has an infinite flag of finite-dimensional
invariant subspaces \eqref{F2} parametrized by $n$. Therefore, formally
it is an exactly solvable operator \cite{turb92}.

From the expression \eqref{H2} we infer the form of the inverse metric:
\begin{equation*}
  \left\|g^{\mu\nu}\right\| =
  \begin{pmatrix}
   p_0x^2 + P_m(y)x + P_{2m}(y)&
   P_1(y)x + P_{m+1}(y)\\P_1(y)x + P_{m+1}(y)&
   P_2(y)
  \end{pmatrix}.
\end{equation*}

In general the system given by the Hamiltonian \eqref{H2} is too
complicated to resolve the constraints \eqref{dA}.
For simplicity we investigate the cases of factorizable and
unfactorizable metric determinants:
\begin{align*}
 \mop{det}\|g^{\mu\nu}\|&\sim
    \bigl(x-\xi_1(y)\bigr)\bigl(x-\xi_2(y)\bigr)
    &\text{and}&&
 \mop{det}\|g^{\mu\nu}\|&\sim
    \left(x-\xi_1(y)\right)^2+\xi_2(y)^2.
\end{align*}
To this end we have to fix the coefficient functions, e.g.:
\begin{align*}
  P_2(y)&=P_2 P_1(y)^2,\\
  P_{m+1}(y)&= \frac 12 P_1(y)\Bigl(P_2P_m(y)+\left(p_0 P_2-1\right)
   \bigl(\xi_1(y)+\xi_2(y)\bigr)\Bigr),\\
  P_{2m}(y)&=\frac{P_2}4
   \left(P_m(y)+\frac{p_0 P_2-1}{P_2}
   \bigl(\xi_1(y)+\xi_2(y)\bigr)\right)^2
   + \frac{p_0 P_2-1}{P_2}\xi_1(y)\xi_2(y)
\end{align*}
for the fist case and
\begin{align*}
  P_2(y)&=P_2 P_1(y)^2,\\
  P_{m+1}(y)&= P_1(y)\left(\frac{P_2}2P_m(y)+\left(p_0 P_2-1\right)
   \xi_1(y)\right),\\
  P_{2m}(y)&=P_2\left(\frac 12P_m(y)
    +\frac{p_0P_2-1}{P_2}\xi_1(y)\right)^2
    + \frac{p_0 P_2-1}{P_2}\left(\xi_1(y)^2+\xi_2(y)^2\right)
\end{align*}
for the second case. The functions $\xi_i(y)$ are polynomials of degrees
no more then $m$.

By shifting and rescaling the variable $y$ we can fix the form of the
linear function $P_1(y)$. There are two different cases. Let us discuss
the first one:
$$
 P_1(y)=y.
$$
In this case there are several solutions of equations \eqref{dA} with
factorizable and unfactorizable metric determinants.
For all of the solutions the coefficient functions $Q_1(y)$, $Q_m(y)$
and $P_m(y)$ are given by
\begin{align}
 Q_1(y)&= Q_1 y,\notag\\\notag
 Q_m(y)&=
  \frac{P_2^2y^2q_m''(y) + Q_1P_2yq_m'(y)
      -\bigl(\left(2 p_0-q_0-2\right)P_2+2 Q_1\bigr)q_m(y)}
       {1-\left(2 p_0-q_0-1\right)P_2-Q_1}
     \\[-3mm]&\label{coef1}\\[-3mm]&
      +\left(p_0-q_0+\frac{Q_1}{P_2}-1\right)
       \bigl(\xi _1(y)+\xi_2(y)\bigr),\notag\\\notag
 P_m(y)&= 2\frac{P_2yq_m'(y)-q_m(y)}{1-\left(2p_0-q_0-1\right) P_2-Q_1}
 +\frac{1-p_0P_2}{P_2}\bigl(\xi_1(y)+\xi_2(y)\bigr).
\end{align}

According to the first solution the polynomials $\xi_i(y)$ are
arbitrary,
$$
 q_m(y)=\left(\frac{1-Q_1}{P_2}-2 p_0+q_0+1\right)\xi_i(y)
$$
and the wave functions have the form
\begin{equation}\label{sol1}
  \psi=g^{-\frac 14}\left|y\right|^\alpha\left|x-\xi_i(y)\right|^\beta
      Pol(x,y),
\end{equation}
where $Pol(x,y)\in\mathcal F_{m,n}$ $i=1$ or $2$ and
\begin{align*}
 \alpha&=\frac{Q_1p_0+p_0-q_0-1}{2\left(p_0P_2-1\right)}-1,&
  \beta&=\frac{q_0 P_2+P_2-Q_1-1}{2\left(p_0 P_2-1\right)}-1.
\end{align*}
The determinant of the inverse metric has the factorizable form
$$
  \mop{det}\|g^{\mu\nu}\|\sim
     y^2\bigl(x-\xi_1(y)\bigr)\bigl(x-\xi_2(y)\bigr).
$$
\begin{figure}[h]
\begin{center}
 \epsfxsize=5.4cm
 \epsfbox{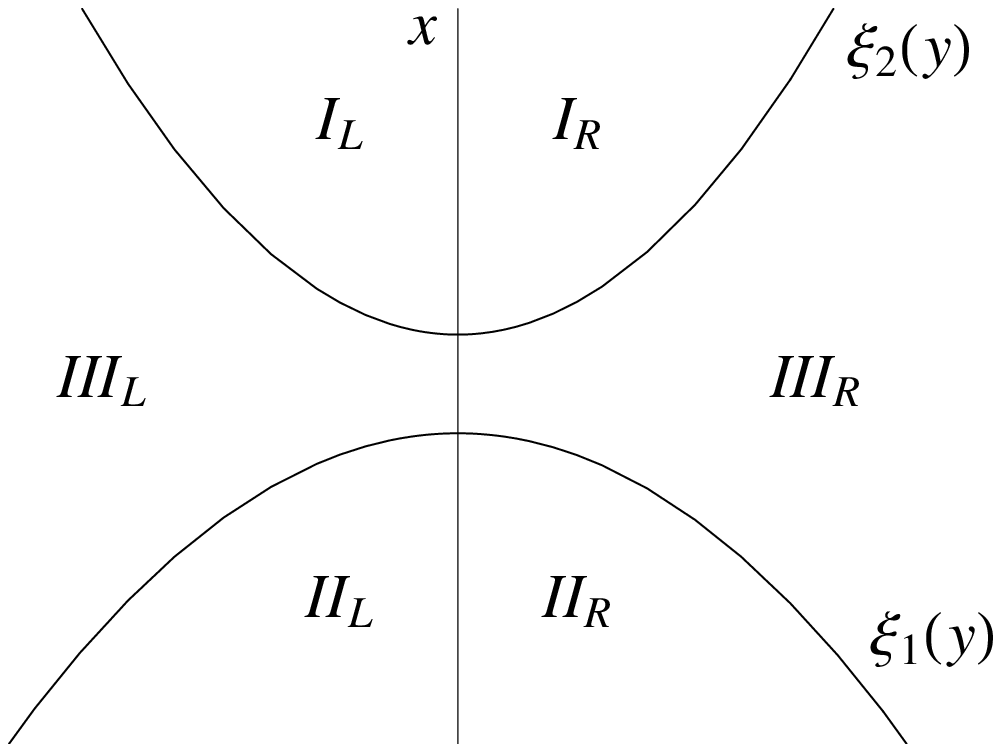}
 \epsfxsize=5.4cm
 \epsfbox{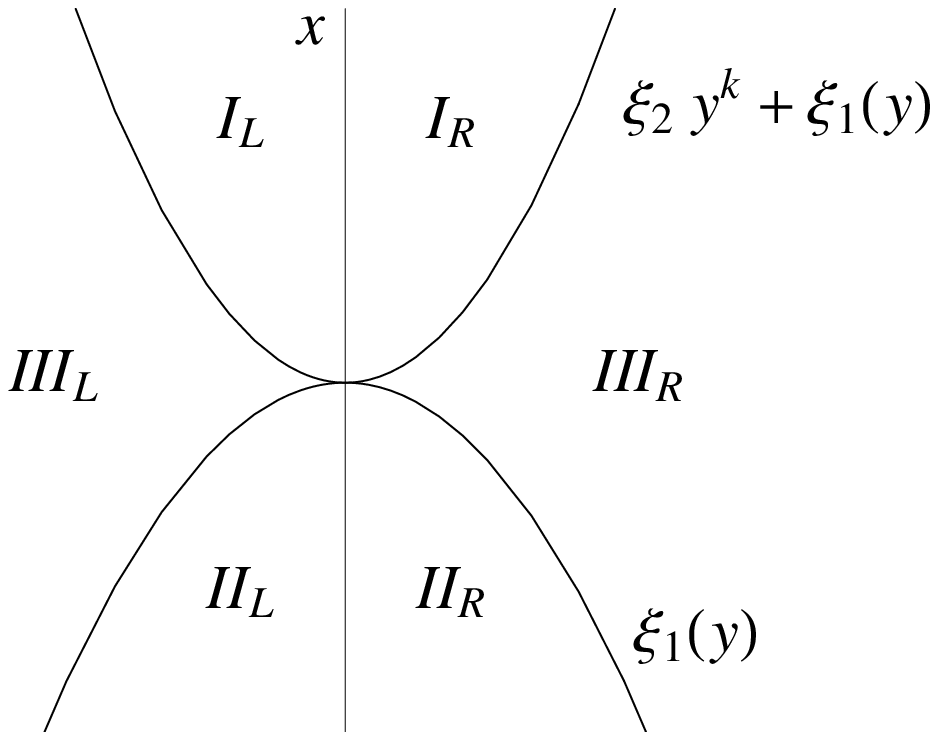}
 \epsfxsize=5.4cm
 \epsfbox{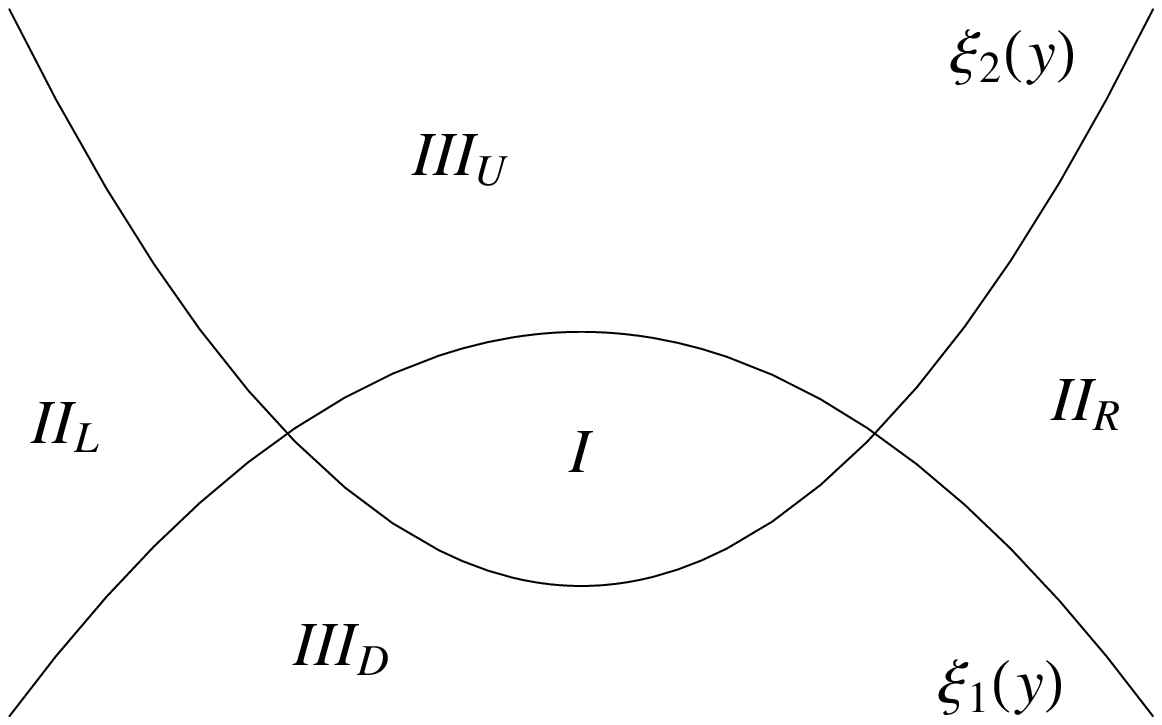}
\end{center}
 \caption{Schematic representations of boundaries.}
\label{pic1}
\end{figure}
Possible configuration of boundaries in this system is schematically
represented on the left plot of figure~\ref{pic1}. From the structure of
the wave functions \eqref{sol1} one can conclude that for the
Hamiltonian to be Hermitian the system should live in the domain $I_L$
($I_R$) with $x\geqslant\xi_2(y)$ and $y\leqslant 0$ ($y\geqslant 0$) or
$II_L$ ($II_R$) with $x\leqslant\xi_1(y)$ and $y\leqslant 0$ ($y
\geqslant 0$). But it is easy to check that for any choice of the
parameters $\alpha>\frac 12$ and $\beta>\frac 12$ the wave functions
\eqref{sol1} are not normalizable. Therefore the system is actually not
quasi-exactly solvable.

In the regions $III_L$ and $III_R$ the wave functions \eqref{sol1} do
not belong to the domain of the Hamiltonian. To consider these domains
it is necessary to discard the hermicity of the Hamiltonian.
Nevertheless in the quasi-exactly solvable sector matrix elements are
real hence corresponding eigenvalues are real or form complex conjugated
pairs. But this pattern is specific to pseudo-Hermitian Hamiltonians
\cite{psH}. Therefore the system in the domains $III_L$ and $III_R$ can
be interesting from this point of view. Moreover some choice of the
parameter $\alpha$ can provide power-like falloff and normalizability of
the wave functions \eqref{sol1} (at least part of them). In this sense
the system is quasi-exactly solvable. Besides if we discard the
hermicity then it is possible to consider another configurations of
boundaries when the corresponding system is quasi-exactly solvable or
even exactly solvable. Such a configuration is represented on the right
plot of the figure~\eqref{pic1}. Indeed in this case the system living
the finite region $I$ is exactly solvable because wave functions are
normalizable for any values of the parameters $\alpha>-\frac 12$ and
$\beta>-\frac 12$.

According to the second solution
\begin{align*}
  \xi_2(y)&=\xi_1(y)+\xi_2y^k,&
   q_m(y)&=\left(\frac{1-Q_1}{P_2}-2 p_0+q_0+1\right)\xi_1(y) + q_my^k
\end{align*}
and the wave functions have the following structure
\begin{equation}\label{sol2}
  \psi=g^{-\frac 14}\left|y\right|^\alpha\left|x-\xi_1(y)\right|^\beta
   \left|x-\xi_1(y)-\xi_2y^k\right|^\gamma Pol(x,y),
\end{equation}
where $Pol(x,y)\in\mathcal F_{m,n}$, $k=P_2^{-1}\in\N$ and
\begin{align*}
 \alpha&=\frac{Q_1p_0+p_0-q_0-1}{2 (p_0 P_2-1)}-1,&
  \beta&=\frac{(q_0P_2+P_2-Q_1-1)\xi_2-P_2 q_m}
              {2(p_0 P_2-1)\xi_2}-1,&
  \gamma&=\frac{P_2 q_m}{2(p_0 P_2-1)\xi_2}.
\end{align*}
The determinant of the inverse metric is factorizable
$$
  \mop{det}\|g^{\mu\nu}\|\sim
     y^2\bigl(x-\xi_1(y)\bigr)\bigl(x-\xi_1(y)-\xi_2y^k\bigr).
$$
Possible configuration of boundaries in this system is schematically
represented on the central plot of the figure~\ref{pic1}. From the
structure of the wave functions \eqref{sol2} one can conclude that for
some choice of the parameters the wave functions are normalizable with
power-like falloff and the system is quasi-exactly solvable in any of
the domains.

According to the third solution the polynomials $\xi_i(y)$ and $q_m(y)$
are not fixed, $Q_1= 1 - P_2\left(2p_0-q_0-1\right)$ and the wave
functions have the following form
$$
 \psi=g^{-\frac 14}\left|y\right|^{\frac{q_0-1}2-p_0}Pol(x,y),
$$
where $Pol(x,y)\in\mathcal F_{m,n}$. Such functions do not belong to the
domain of the Hamiltonian or are not normalizable for any choice of the
parameters. Therefore the corresponding system is not quasi-exactly
solvable. Nevertheless if one discards the hermicity of the Hamiltonian
then for the situation represented on the right plot of
figure~\ref{pic1} the system can be quasi-exactly solvable (in the
regions $II_L$ and $II_R$) or even exactly solvable (in the region $I$)
for some choice of the parameters $q_0$ and $p_0$.

Now we pass to the metric with the unfactorizable determinant, i.e.
it has no roots in the variable $x$. In this case there is the following
nontrivial solution:
\begin{align*}
 \xi_2(y)&=\xi_1(y)+\xi_2y^k,&
   q_m(y)&=q_m\xi_2y^k-\left(Q_1k-k+2p_0-q_0-1\right)\xi_1(y)
\end{align*}
and the wave functions have the following structure
\begin{equation}\label{sol3}
 \psi=g^{-\frac 14}\left|y\right|^\alpha
 \left(\left(x-\xi_1(y)\right)^2+\xi_2^2y^{2k}\right)^\beta
  e^{\gamma\arctan\frac{\xi_2y^k}{x-\xi_1(y)}}Pol(x,y),
\end{equation}
where $Pol(x,y)\in\mathcal F_{m,n}$, $k=P_2^{-1}\in\N$ and
\begin{align*}
 \alpha&=\frac{k\left(Q_1p_0+p_0-q_0-1\right)}
              {2\left(p_0-k\right)}-1,&
  \beta&=\frac{Q_1k+k-q_0-1}{4\left(k-p_0\right)}-\frac 12,&
 \gamma&=\frac{q_m}{2 \left(k-p_0\right)}.
\end{align*}
The determinant of inverse metric has the form
$$
  \mop{det}\|g^{\mu\nu}\|\sim
    y^2\left(\left(x-\xi_1(y)\right)^2+\xi_2^2y^{2k}\right).
$$
The determinant vanishes only at the line $y=0$, therefore the system
can live on the left or right semiplane of $Oyx$. One can see that
negative values of the parameter $\beta$ with large enough modulus can
provide power-like falloff and normalizability of a finite number of the
wave functions \eqref{sol3}. Therefore the system is quasi-exactly
solvable.

Now let us consider the second possibility of fixing the linear function
$P_1(y)$:
$$
 P_1(y)=P_1.
$$
In this case for all of the solutions the coefficient functions
$Q_1(y)$, $Q_m(y)$ and $P_m(y)$ are given by
\begin{align}
 Q_1(y)&= P_1P_2Q_1 y,\notag\\\notag
 Q_m(y)&=
  \frac{P_1^2P_2^2q_m''(y) + P_1P_2^2Q_1q_m'(y)
      -P_2\left(2 p_0-q_0+2Q_1\right)q_m(y)}
       {1-\left(2 p_0-q_0+Q_1\right)P_2}
     \\[-3mm]&\label{coef2}\\[-3mm]&
      +\left(p_0-q_0+Q_1\right)\bigl(\xi _1(y)+\xi_2(y)\bigr),
       \notag\\\notag
 P_m(y)&= 2\frac{P_1P_2q_m'(y)-q_m(y)}{1-\left(2p_0-q_0+Q_1\right)P_2}
     +\frac{1-p_0P_2}{P_2}\bigl(\xi_1(y)+\xi_2(y)\bigr).
\end{align}

According to the first solution the polynomials $\xi_i(y)$ are
arbitrary,
$$
 q_m(y)= \left(\frac 1{P_2} - 2p_0 + q_0 - Q_1\right)\xi_i(y)
$$
and the wave functions have the form
\begin{equation}\label{sol21}
  \psi=g^{-\frac 14}\left|x-\xi_i(y)\right|^\alpha e^{-\beta y}Pol(x,y),
\end{equation}
where $Pol(x,y)\in\mathcal F_{m,n}$, $i=1$ or $2$ and
\begin{align*}
 \alpha&=\frac{P_2\left(q_0-Q_1\right)-1}
              {2\left(p_0P_2-1\right)}-1,&
 \beta&=\frac{q_0-p_0 \left(P_2 Q_1+1\right)}
    {2 P_1\left(p_0P_2-1\right)}.
\end{align*}
The determinant of the inverse metric has the form
\begin{equation}\label{det21}
  \mop{det}\|g^{\mu\nu}\|\sim
    \bigl(x-\xi_1(y)\bigr)\bigl(x-\xi_2(y)\bigr).
\end{equation}
For the Hamiltonian to be Hermitian the functions $\xi_i(y)$ have to
obey the inequality $\xi_1(y)\geqslant\xi_2(y)$ or
$\xi_1(y)\leqslant\xi_2(y)$. Such a configuration is represented on the
left side of figure~\ref{pic1}. One can check that for any choice of the
parameters $\alpha$ and $\beta$ the wave functions \eqref{sol21} are not
normalizable. Therefore the system is actually not quasi-exactly
solvable. Nevertheless if non-Hermitian Hamiltonians are allowed then
for configurations of boundaries represented on the right plot of
figure~\ref{pic1} the systems are exactly solvable in the regions $I$,
$II_L$ and $II_R$.

Also there is a solution with the determinant of the form \eqref{det21}
with arbitrary polynomials $\xi_i(y)$, $q_m(y)$ and wave functions with
the structure $\psi\sim g^{-1/4}e^{-\beta y}Pol(x,y)$.
In this case hermicity of the Hamiltonian or normalizability cannot be
provided. Therefore the system is not quasi-exactly solvable. But if we
discard the hermicity of the Hamiltonian then for the configuration
represented on the right plot of figure~\ref{pic1} the system is exactly
solvable if it lives in the domains $I$, $II_L$ and $II_R$.

\section{Conclusion}
\label{end}

In this paper we have considered quantum-mechanical systems in domains
with boundaries. Such a situation is usual for the Lie-algebraic
approach \cite{qes, shifman} to construction of multidimensional
quasi-exactly solvable Hamiltonians. The hermicity of the operators
prescribes the specific behaviour of wave functions which belong to
domains of Hamiltonians. By an example it was shown that not all of
known quasi-exactly solvable systems are in fact quasi-exactly solvable
even if corresponding wave functions are normalizable.

Besides we have constructed new two-dimensional families of formally
exactly solvable systems. Application of the restrictions the behaviour
on wave functions in a domain with boundaries leads to the conclusion
that the constructed systems are quasi-exactly solvable at best.

If one discards the hermicity then the behaviour of the wave functions
is governed only by the normalizability. It is worth noting that the
corresponding matrix elements are real. In the quasi-exactly solvable
sector the corresponding matrix is finite dimensional therefore
quasi-exactly solvable eigenvalues are defined by an algebraic equation
with real coefficients. This means that the eigenvalues are represented
by real values and complex conjugated pairs of numbers but this pattern
corresponds exactly to the spectrum of a pseudo-Hermitian Hamiltonian
\cite{psH}. Therefore the formally quasi-exactly solvable and exactly
solvable systems can be interesting in this context. Of course, the
exact relationship between such quasi-exactly solvable operators and
pseudo-Hermitian Hamiltonians requires more detailed investigation.


\end{document}